\documentclass{aa}  

\usepackage{graphicx}
\usepackage{txfonts}
\pdfoutput=1
\begin{document} 

\title{The young stellar population in NGC 247}
\subtitle{Main properties and hierarchical clustering}

\author{
M. J. Rodríguez\inst{1}\thanks{jimenaro@fcaglp.unlp.edu.ar},
G. Baume\inst{1,2} and 
C. Feinstein\inst{1,2}
}
\institute{
Instituto de Astrofísica de La Plata (CONICET-UNLP), Paseo del bosque S/N, 
La Plata (B1900FWA), Argentina, \and 
Facultad de Ciencias Astronómicas y Geofísicas - Universidad Nacional de La Plata, Paseo del bosque S/N, La Plata (B1900FWA), Argentina}

\date{Accepted XXX. Received YYY; in original form ZZZ}

 
\abstract
{}
{The aim of this work is to investigate the characteristics of the young stellar population in the spiral galaxy NGC~247. In particular, we focused our attention in its hierarchical clustering distributions and the properties of the smallest groups.}
{We used multiband Hubble Space Telescope (HST) data from three fields covering more than half of NGC~247 to select the young population. Then, through the  path linkage criterion (PLC), we found compact young star groups, and estimated their fundamental parameters, such as their stellar densities, sizes, amount of members and luminosity function (LF) slopes. We also performed a fractal analysis to determinate the clustering properties of this population. We build a stellar density map and the corresponding dendrograms corresponding to the galactic young population to detect large structures and draw their main characteristics.}
{We detected 339 young star groups, for which we computed a mean radius of $\sim$ 60~pc and a maximum in the size distribution between 30 and 70~pc. We also obtained LF slopes with a bimodal distribution showing peaks at $\sim$ 0.1 and $\sim$ 0.2. We identified several candidates to HII regions which follow an excellent spatial correlation with the young groups found by the PLC. We observed that the young population are hierarchically organized, where the smaller and denser structures are within larger and less dense ones. We noticed that all these groups presented a fractal subclustering, following the hierarchical distribution observed in the corresponding stellar density map. For the large young structures observed in this map, we obtained a fractal dimension of $\sim$ 1.6-1.8 using the perimeter-area relation and the cumulative size distribution. These values are consistent with a scenario of hierarchical star formation.}
{}

\maketitle
%

\section{Introduction}

Most massive stars do not form in isolation, they do in embedded cluster \citep{2003ARA&A..41...57L}. Therefore, the young star groups constitute the fundamental blocks of star formation in galaxies. Their study is essential to understand the star formation processes in the host galaxy as well as its recent history. 

Different young stellar groups are found at diverse scales, from young compact open clusters, to important stellar complexes of hundreds of parsecs, going through OB associations and different stellar aggregates. All these young structures are not independent of each other, they are linked in a hierarchical way, in which the most compact and dense structures are within larger and less dense ones \citep[e.g.][]{1996ApJ...466..802E, 1998MNRAS.299..588E}. Thus, the OB associations host stellar subgroups and open clusters \citep{1964ARA&A...2..213B}, and they are at the same time part of a larger and looser stellar aggregate \citep{2018MNRAS.479..961R,2016A&A...594A..34R}. It is usually accepted  that this hierarchical behavior is inherit from their parent molecular clouds \citep[e.g.][]{1999AJ....117..764E}. The interstellar molecular gas presents a self-similar distribution, known as fractal structure, which seems to be driven mainly by turbulence and self gravity \citep{1996ApJ...466..802E,1999ApJ...527..266E}.

In this way, the star formation proceeds continuously along several scales \citep{2018PASP..130g2001G}. Thus, it seems arbitrary to set a bound into the different groups (clusters, OB associations, aggregates, complexes, supercomplexes) based only on their morphology. However, such separation is useful to study the young stellar population at different scales and be able to establish their characteristics.
 
In this work we present a study of the young stellar population in the Sculptor Group spiral galaxy NGC~247. This galaxy is catalogued as a SAB(s)~d~D \citep{2013AJ....145..137K} and is located near the South Galactic Pole ($l = 113.95^{\circ}$; $b = -83.56^{\circ}$) having a high inclination angle ($i = 74^{\circ}$; \citealt{2000A&A...356L..49S}) and an estimated distance of 3.6 Mpc \citep{2008AJ....136.1770G}. \\

Several studies have been carried out in this galaxy using ground based observations. \citet{2006ApJ...641..822D} reported, using deep visible and near-infrared images, the presence of an extended stellar disk with star formation, and derived a star formation rate (SFR) of 0.1 M$\sun$~yr$^{-1}$ in the last 16 Myr. \citet{1990AJ....100..641C} studied the properties of this galaxy in HI using VLA observations. They found a relatively compact HI envelope compared to other systems. On the other hand, \citet{1996AJ....111.2265F} noted that the H$_{\alpha}$ emission was mainly from two spiral arms, and they found a small total H$_{\alpha}$ flux compared to other spirals and irregulars galaxies.  

In spite of its proximity, detailed studies have not been done about its stellar population. There are not catalogs of stellar associations, and there are only records of a few identified clusters in the literature. \citet{1999A&AS..139..393L} identified three massive star clusters in NGC~247, but only one of them was associated with a blue, and therefore young, population. \citet{2004AJ....127.2674O} detected three globular clusters, and \citet{2012ApJ...758...85T} using visible HST and x-ray Chandra observations, reported the existence of a young stellar association close to an ultraluminous X-ray source (ULX).

In this paper, we took advantage of the excellent quality data obtained with the Hubble Space Telescope (HST), to study for the first time the young stellar population of NGC~247.
We focused mainly on the young star groups obtaining their principal characteristics, and on the study of the hierarchical organization observed in the young population. 

This work is organized as follow. In Sect.~\ref{S_data} we describe the observations and photometry data used in this research. In Sect.~\ref{S_search} are described the methods used for the identification of the young stellar population and for the search of stellar groups inside it. In Sect.~\ref{parameters_sect} we describe the principals characteristics of the detected groups. The detection of larger young structures is described in Sect.~\ref{largerstructures}. A morphological analysis of the detected groups and structures is presented in Sect.~\ref{morphology}. Finally, in Sec.~\ref{conclusions} we present our conclusions. 


\section{Data}
\label{S_data}

\subsection{Observations}

\begin{figure}
\resizebox{\hsize}{!}{\includegraphics{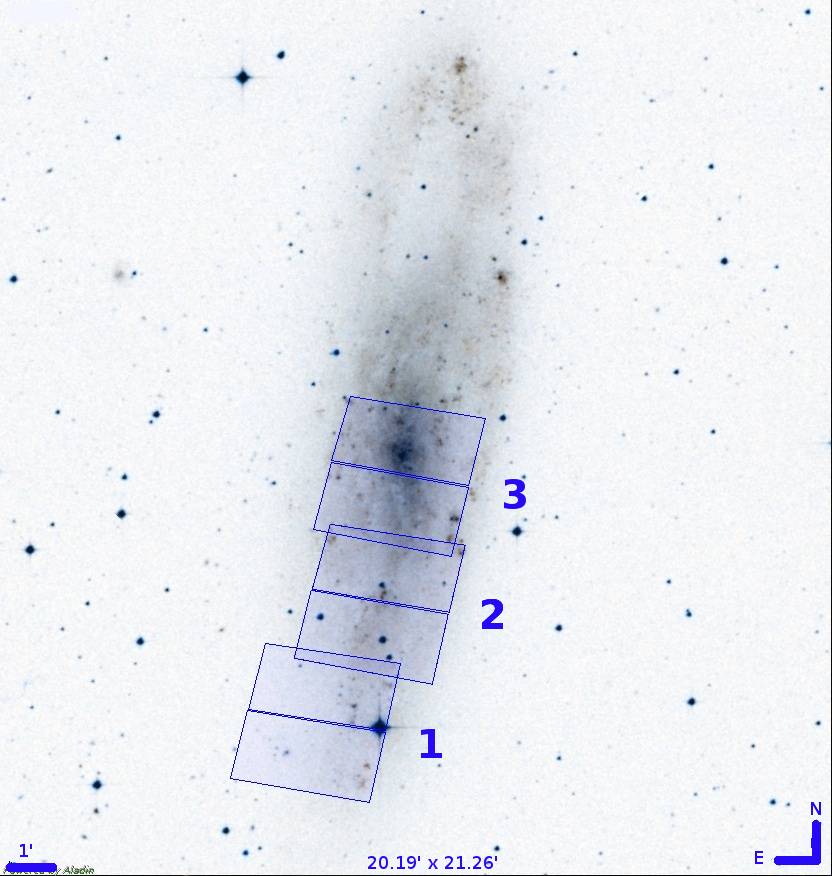}}
\caption{Distribution of the different HST/ACS fields (rectangles) used in this work overlaid on a Digitized Sky Survey (DSS) image of NGC~247. The North is up and East is left, the image cover $20.19'\times 21.26'$.}
\label{campos}
\end{figure}

In this work we used images obtained with the Wide Field Camera (WFC) of the Advanced Camera for Surveys (ACS). They correspond to three fields in NGC~247, which cover a total length of $\sim$10' encompassing the southern half of the galaxy (see Fig. \ref{campos}). Each field was observed using the broadband HST filters $F475W, F606W$ and $F814W$. 

The observations were carried out in September 2006 during the HST Cycle 15, as part of the program GO-10915 (PI: J. Dalcanton). The observations details are listed in Table \ref{T_observations}. The WFC has a mosaic of two CCD detectors with a field of view of $3.3'~\times~3.3'$ and a scale of $0.049"$/pixel. The observation described above were acquired from the Hubble Legacy Archive\footnote{http://hla.stsci.edu/}.

\begin{table}
\caption{Observations details}             
\label{T_observations}      
\centering          
\begin{tabular}{c c c c c}     
\hline\hline       
Field & Band   & N$_{exp}$  &  t$_{exp}$ [sec] & Date \\
\hline
1     & F475W  & 3          & 2253             & 09-22-2006  \\
1     & F606W  & 3          & 2280             & 09-22-2006  \\
1     & F814W  & 3          & 2250             & 09-22-2006  \\
2     & F475W  & 2          & 1480             & 09-20-2006  \\
2     & F606W  & 2          & 1507             & 09-20-2006  \\
2     & F814W  & 2          & 1534             & 09-20-2006  \\
3     & F475W  & 2          & 1480             & 09-21-2006  \\
3     & F606W  & 2          & 1507             & 09-21-2006  \\
3     & F814W  & 2          & 1534             & 09-21-2006  \\
\hline              
\end{tabular}
\end{table}

\subsection{Photometry}

The photometry data used in this work was obtained from the MAST data base of the Space Telescope Institute (STScI)\footnote{https://archive.stsci.edu/}. They correspond
to the \textit{“star files”} of the \textit{ACS Nearby Galaxy Survey} (ANGST).

The point spread function (PSF) photometry was carried out by \citet{2008ASSP....5..115D}, using the software {\ttfamily DOLPHOT}\footnote{http://americano.dolphinsim.com/dolphot/} adapted for the ACS camera \citep{2000PASP..112.1383D}. The files mentioned above contain the photometry of all objects classified as stars with good signal-to-noise values ($S/N~>~4$) and data flag~$<$~8. 

\begin{figure}
\resizebox{\hsize}{!}{\includegraphics{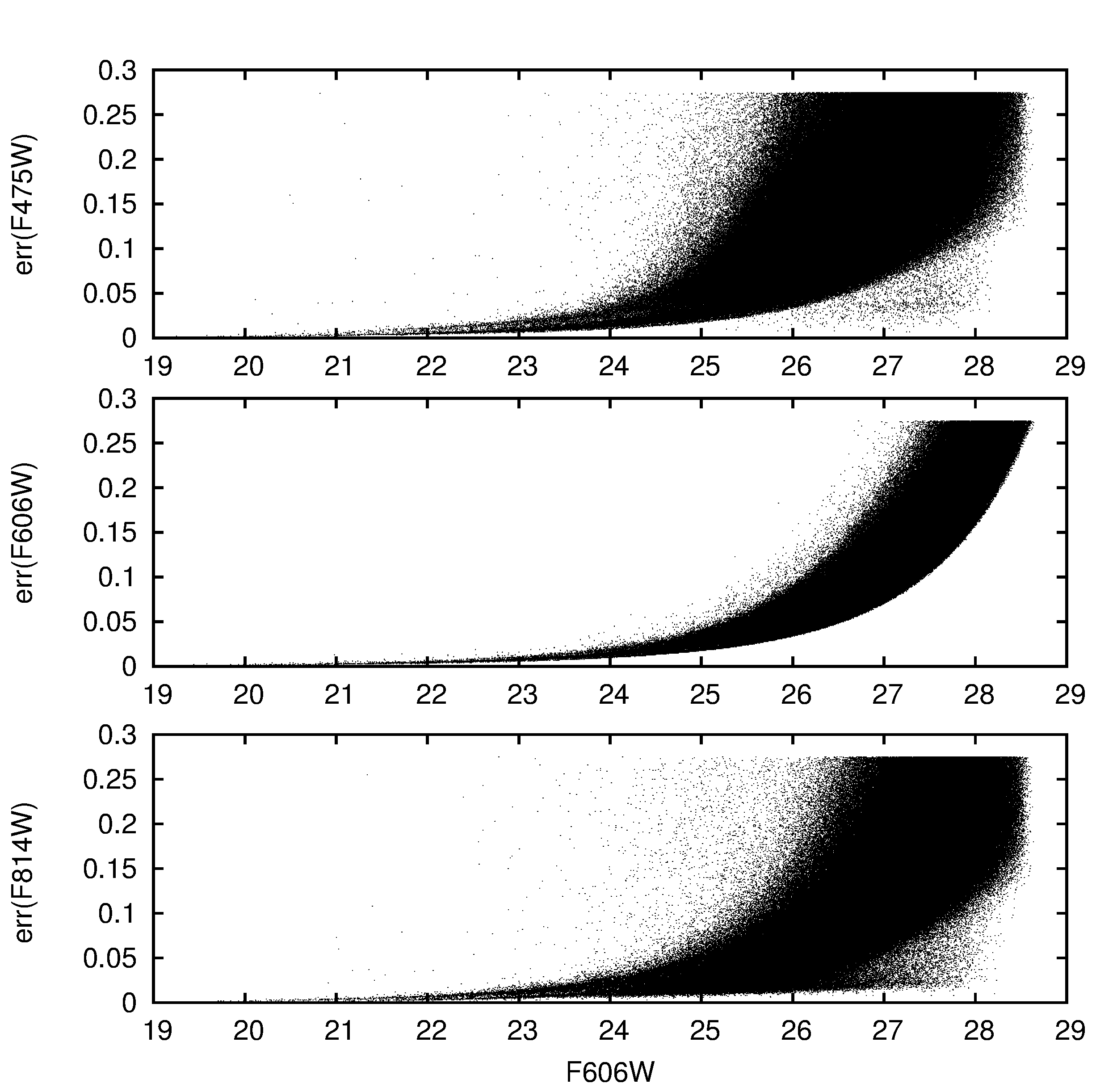}}
\caption{Photometric errors in the different bands}
\label{errors}
\end{figure}

\begin{figure}
\resizebox{\hsize}{!}{\includegraphics{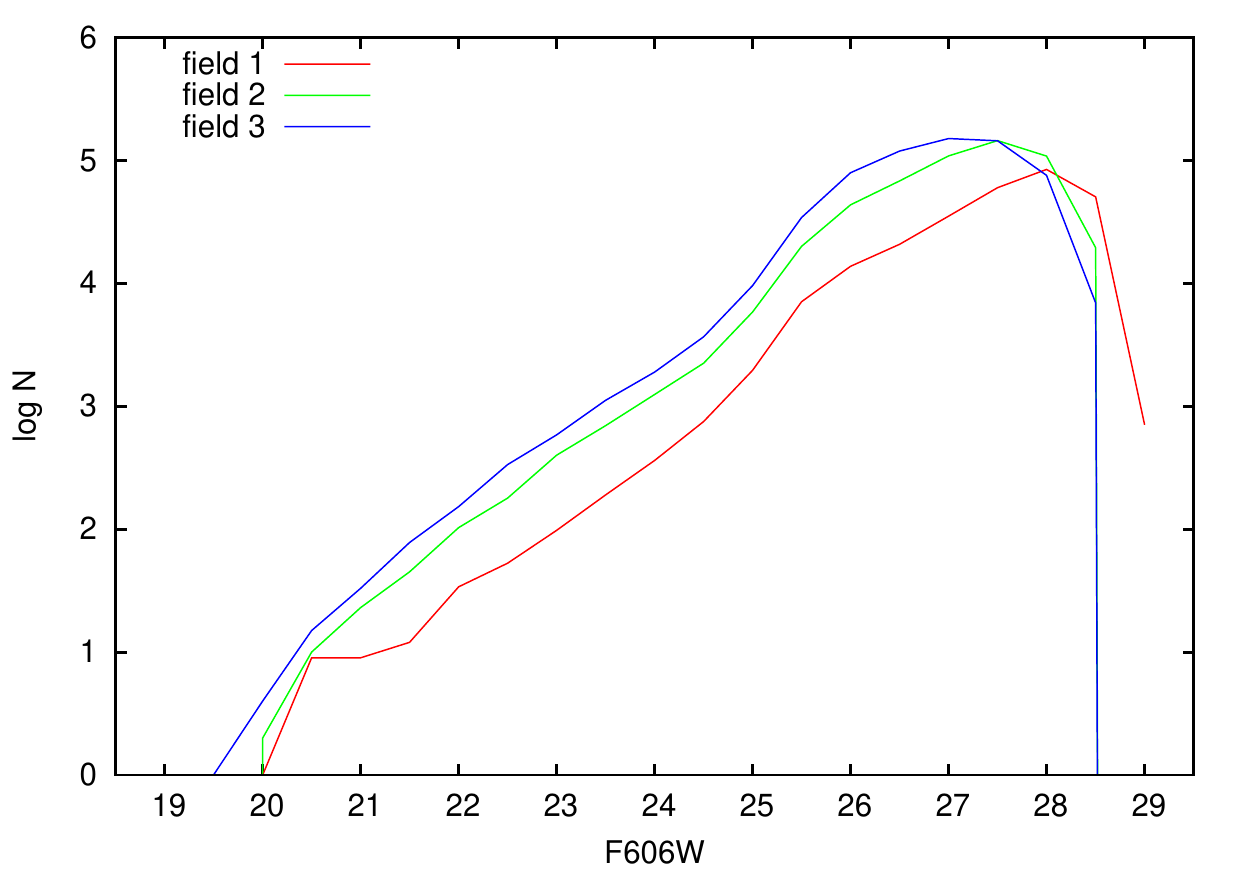}}
\caption{Luminosity Function for the three studied fields}
\label{LFtotal}
\end{figure}

In Fig.~\ref{errors} we show the photometric errors in the different bands corresponding to the objects classified as stars in the three fields. We built the $F606W$ luminosity function (LF) for the three fields (see Fig.~\ref{LFtotal}) in order to evaluate the completeness of the data. We can see in Fig.~\ref{LFtotal} that the number of stars per bin starts to decrease at $F606W~\sim~27.5$. Therefore we considered that the sample is complete up to this value. 

\subsection{Photometric correlation of tables}
\label{catalog}

The ANGST provides photometric tables with information of only two bands. To join the three magnitudes in a single table for field, we used the code STILTS\footnote{http://www.star.bris.ac.uk/$\sim$mbt/stilts/} to perform a cross-correlation (with logical "OR") between tables with $F606W-F475W$ bands and those with $F814W-F606W$ bands, obtaining three photometric tables, one for each field.

As we can note in Fig.~\ref{campos}, the adjacent fields slightly overlap each other. We used STILTS to merge the information of the three fields in one single table with approximately 1.3$\times$10$^6$ objects. In this procedure we used only the brightest objects ($F606W~<~24$). We first compared the astrometric information with objects of Frame 3 in common with GAIA DR2 and we found a small systematic difference that was corrected. Then we compared the astrometric and photometric data of the objects in the overlapped regions and we found small differences in the astrometry but not important differences ($\Delta mag \sim 0.01$) in their photometry values. Finally, we adopted the central part of NGC~247 (Frame 3) as a reference and we carried out coordinates of Frames 1 and 2 to the former. For objects in overlapped regions, we took, after the calibration, the average of data between the stars in each field. The computed astrometric differences are presented in Table~\ref{T_shift}.

\subsection{Stellar blend and foreground contamination}

Star blends can be found in the case that two (or more) stars have not the same center but are close enough to be considered as one object. The software used to measure the photometry of the stars (DOLPHOT) gives as a result for each object the sharp, the crowd, the S/N and the $\chi^2$ of the PSF fitting. These parameters are a good estimator of the quality of the measurement and a reference for detecting blended stars. The most problematic case is when two stars that could be centered at the same position of a pixel and mistakenly detected by the photometry software as only one star. We could not identify these stars by photometry or other mean, but we can modeled how severe or not the effect of this overlap could be. 

An estimation of the maximum level of blending can be calculated by measuring the maximum star density in the galaxy and assuming that this stellar density is constant over the whole object. So, we can model the blending as considering that the stars are distributed with a Poisson distribution. This is equivalent to the procedure of \cite{Kiss} that is this deeply discussed for this kind of observations in \citet{Feinstein}. So, doing this we get an upper limit of probability of blends for the crowded regions, that is larger than the probability of the less populated places. By this procedure we got that the probability of losing stars by blends is less than 2\% in more crowded locations. So, we do not consider blending as an important issue in this work. 

One possible source of data contamination could be the galactic stars in the field of view of NGC~247. However, this galaxy has galactic coordinates $l = 113.95^{\circ}$ $b = -83.56^{\circ}$ which locates the galaxy close to the South pole of the Milky Way and far away from the disk and the galactic center. This means that the probability of contamination of our sample by blue galactic stars is negligible. In order to recheck this possible contamination we ran several simulations with the Besançon model of the galaxy\footnote{https://model.obs-besancon.fr/} and also with the Trilegal\footnote{http://stev.oapd.inaf.it/cgi-bin/trilegal} \citep{Trigal} model and in both cases we found that in very few models we got at least one blue Milky Way's star for an area equivalent to the observed one. So, we do not consider the galactic contamination as a problem for the analysis of young blue stars

\begin{table}
\caption{Astrometric corrections applied ACS/HST data} 
\label{T_shift}      
\centering          
\begin{tabular}{cccc} 
\hline\hline                             
Fields & $\Delta \alpha$["] & $\Delta \delta$["] & $N$ \\
\hline
3-GAIA &  0.586 $\pm$ 0.071 & -0.030 $\pm$ 0.025 & 63 \\
2-3    & -0.172 $\pm$ 0.058 &  0.082 $\pm$ 0.040 & 76 \\
1-2    &  0.312 $\pm$ 0.002 &  0.110 $\pm$ 0.004 & 12 \\
\hline
\end{tabular}
\begin{minipage}[]{0.4\textwidth}
{\bf Note:} Values are given as $mode \pm stdev$ and $N$ indicate the amount of objects used in each case
\end{minipage}
\end{table}

\section{Search of young stellar groups}
\label{S_search}
\subsection{Identification of the different star populations}
\label{populations}

\begin{figure}
\resizebox{\hsize}{!}{\includegraphics{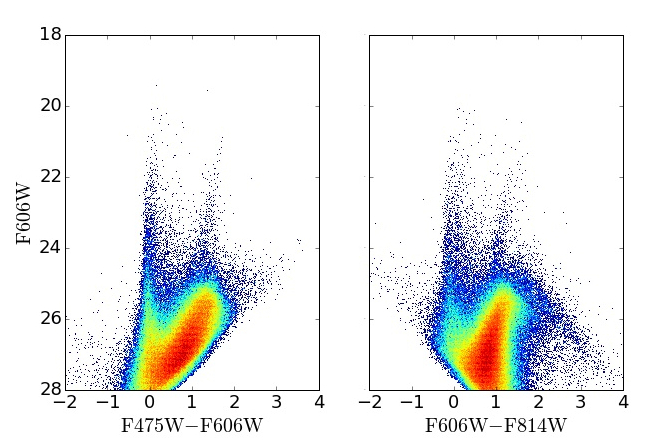}}
\resizebox{\hsize}{!}{\includegraphics{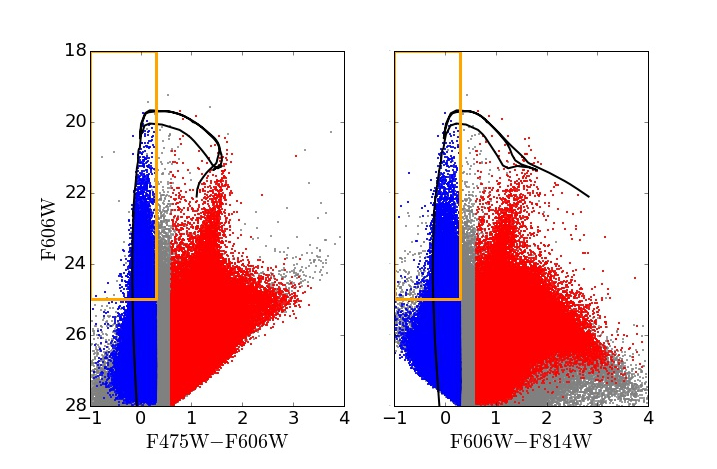}}
\caption{{\it Top panel:} Combination of Hess diagram and CMDs for the detected objects in the three observed fields. The different colors indicates densities, ranging from blue (lower densities) to red (higher densities). We show the diagrams correspond to the color indexes $F475W-F606W$ and $F606W-F814W$ at left and right respectively.
{\it Bottom panel:} The same CMDs that in the top panel are showed, but here we separate the blue and red population (with blue and red color respectively). The orange rectangle encloses the bright blue stars, we use this selection to search young stellar groups. In this figure we overlapped the PARSEC version 1.2S isochrone \citep{2014MNRAS.445.4287T} corresponding to 10$^7$ years and metallicity Z = 0.0152. This model was displaced adopting a distance modulus of $(V_0 - M_V )$ = 27.78 \citep{2008AJ....136.1770G}, a normal reddening law ($R = A_V /E(B-V)$ = 3.1) and a value for $E(B-V)$ = 0.015 corresponding to the foreground reddening toward NGC 247 \citep{2011ApJ...737..103S}.}
\label{hess-cmd}
\end{figure}

In the top panel of Fig.~\ref{hess-cmd} we show a combination of Hess and color magnitude diagram (CMD) for the detected objects in the studied region of NGC~247, where the different colors indicate densities, ranging from blue (lower densities) to red (higher densities). Here is possible to note two different populations. One at the left showing a sequence up to reach magnitudes close to $F606W\sim$ 19, which is associated with a young population, and a second sequence at the right which correspond to a more evolve population. Taking into account this, we adopted the following criteria to separate the different stellar populations:

\begin{itemize}
\item Bright stars: $F606W < 25$
\item Blue stars: \\ $(F475W-F606W) < 0.3$ and $(F606W-F814W) < 0.3$
\item Red stars: \\ $(F475W-F606W) > 0.6$ and $(F606W-F814W) > 0.6$
\end{itemize}


In the bottom panel of Fig~\ref{hess-cmd} we indicate with blue and red color, the blue and red population respectively, derived from the above criteria. The rectangle encloses the blue bright stars, which will be used to identify young stellar groups with the PLC method as describe below. In this figure we overlapped the PARSEC version 1.2S evolutionary model \citep{2014MNRAS.445.4287T} corresponding to 10$^7$ years and metallicity Z = 0.0152, to make a comparison with a young stellar population. This isochrone was displaced adopting a distance modulus of $(V_0 - M_V )$ = 27.78 \citep{2008AJ....136.1770G}, a normal reddening law ($R = A_V /E(B-V)$ = 3.1) and a value for $E(B-V)$ = 0.015 corresponding to the foreground reddening toward NGC 247 \citep{2011ApJ...737..103S}. The following coefficients were used to transform these values to the HST system $A_{F475W}/A_V$~=~1.192, $A_{F606W}/A_V$=~0.923 and $A_{F814W}/A_V$=~0.605\footnote{http://stev.oapd.inaf.it/cgi-bin/cmd} \citep{1994ApJ...437..262O}.

In both panels is possible to note that the main sequence population is very well delineated by a thin line of blue stars. This fact suggests that differential extinction in this galaxy would be negligible. In order to check this, we compared the position in the CMD diagrams of all the bright stars that were not intrinsically red ($F475W-F606W<1$) with the theoretical model of 10$^7$ years, showed in the bottom panel of Fig. \ref{hess-cmd}. Following the procedure used in \citet{2018MNRAS.479..961R}, we estimated individual absorption values for these stars. We found small values of $A_V$, with a mode of 0.1 mag and a mean value of 0.2 mag in their distribution. According to these results, we considered that the effects of the differential extinction in this galaxy is negligible for the detection of young groups.

\begin{figure}
\resizebox{\hsize}{!}{\includegraphics{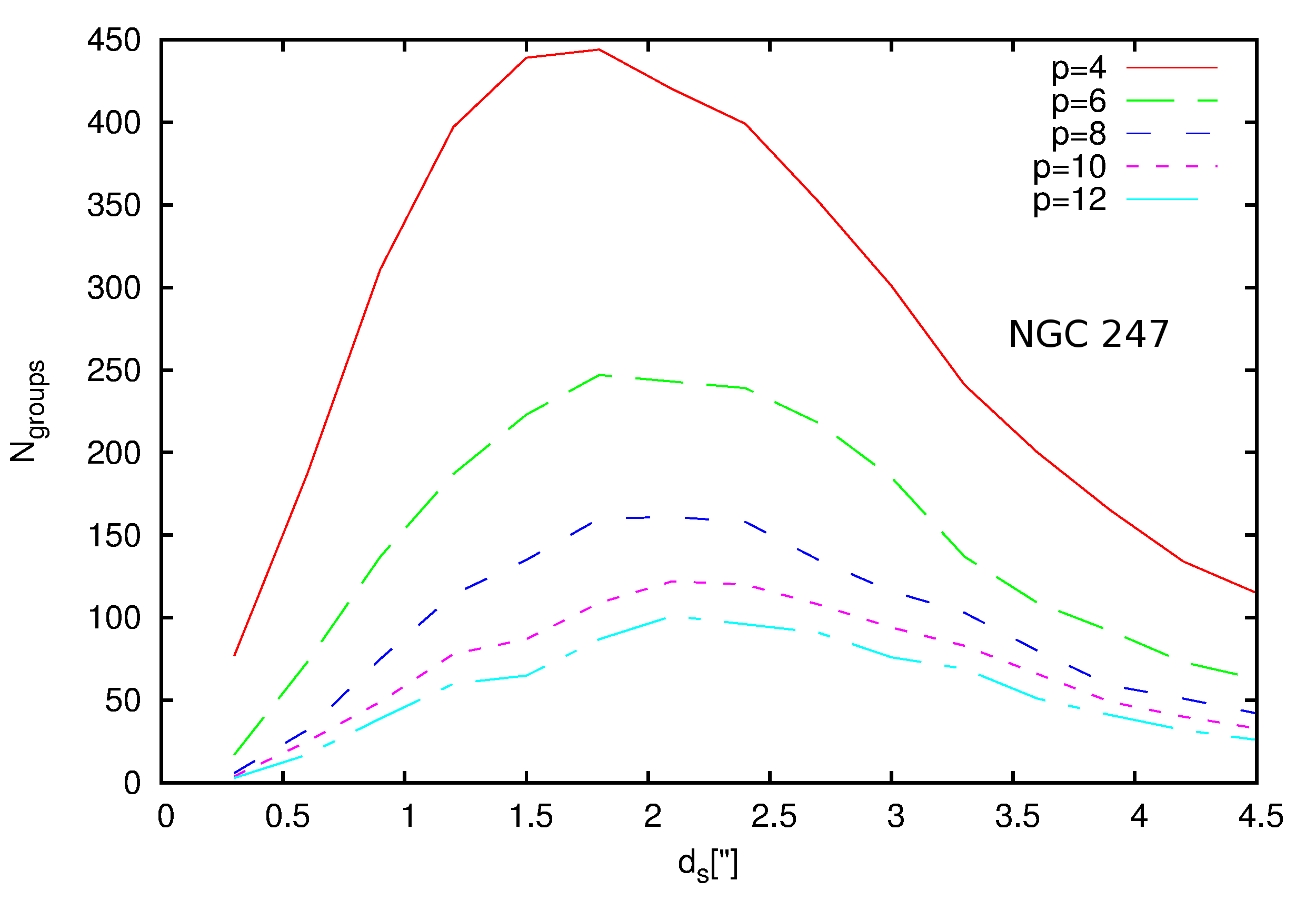}}
\caption{Number of stellar groups detected with the PLC as a function of the search radius $d_s$ for different values of $p$ }
\label{plc_ds}
\end{figure}

\subsection{Searching method}
\label{plc_sect}

We used the Path linkage criterion \citep[PLC,][]{1991A&A...244...69B} over the bright blue population to detect young stellar groups in the galaxy. The applied procedure was explained in our previous works \citep{2018MNRAS.479..961R,2016A&A...594A..34R}. The PLC is an agglomerative friend of friend algorithm, that links stars of the bright blue sample using a search radius $d_s$, and considers a minimum number of stars ($p$) to define an stellar association. In the Fig.~\ref{plc_ds} we show the number of stellar groups detected as a function of $d_s$ for different $p$ values. Here we can see that independent of the value of $p$ the maximum of the distribution falls around 2~arcsec. This maximum corresponds to the optimal value of $d_s$ according to \citet{1991A&A...244...69B}. We decided to adopt a range of $d_s$ values between 0.3-2.0~arcsec (5-35~pc), and in this way we ensure that we are not losing the smallest groups. The PLC method starts the search with $d_s$~=~0.3~arcsec. Then, the stars belonging to the detected groups are deleted from the sample and the PLC runs again increasing the value $d_s$ in a step of 0.4~arcsec. This procedure is repeated until $d_s$ reaches the value of 2~arcsec. We also considered from the results showed in Fig.~\ref{plc_ds} that $p$~=~8 stars is an adequate value. Using this method we detected 339 young stellar groups. 

It is important to note that although the PLC method is devised to detect stellar associations, we can not rule out that some open clusters or  stellar aggregates larger than a typical association are detected. For this reason we named "stellar groups" to refer to all the groups detected, although it is expected that the most of them be stellar associations. 

\begin{table*}
\tiny
\caption{Main parameters of young stellar groups in NGC~247}
\label{tabla1}
\centering 
\begin{tabular}{cccccccccccc} 
\hline\hline
(1) & (2) & (3) & (4) & (5) & (6) & (7) & (8) & (9) & (10) & (11) & (12) \\
Name	& $\alpha_{2000}$ & d$_{GG}~[Kpc]$ & $r~[pc]$ & $N_g$	 & $N_d$	& $F606W_{min}$	& $\rho~[obj./pc^3]$ & $\Gamma$     & $\overline{m}$     & $\overline{s}$     & $Q$   \\
        & $\delta_{2000}$ &                &          & $N_{bg}$ & $N_{bd}$	&               &       	         & $e_{\Gamma}$ & $e_{\overline{m}}$ & $e_{\overline{s}}$ & $e_Q$ \\
\hline
AS001	& ~00:47:13.4	& 10.28	& 11.87	& ~22	& ~22	& 21.24	& 0.00129	& 0.05	& 0.74	& 0.93	& 0.79 \\
     	& -20:51:40.5	&      	&      	& ~~9	& ~~9	&      	&        	& 0.04	& 0.08	& 0.08	& 0.09 \\
AS002	& ~00:47:08.8	& ~6.66	& 11.52	& ~18	& ~14	& 21.27	& 0.00156	& 0.02	& 0.72	& 0.90	& 0.80 \\
     	& -20:50:54.7	&      	&      	& ~10	& ~10	&      	&        	& 0.07	& 0.09	& 0.09	& 0.08 \\
AS003	& ~00:47:08.7	& ~0.99	& 10.82	& ~25	& ~12	& 22.19	& 0.00170	& 0.07	& 0.83	& 1.09	& 0.76 \\
     	& -20:44:45.8	&      	&      	& ~~9	& ~~9	&      	&        	& 0.25	& 0.09	& 0.09	& 0.10 \\
AS004	& ~00:47:10.1	& ~4.96	& 19.20	& ~66	& ~49	& 21.02	& 0.00115	& 0.05	& 0.62	& 0.92	& 0.67 \\
     	& -20:48:55.7	&      	&      	& ~34	& ~34	&      	&        	& 0.05	& 0.04	& 0.04	& 0.06 \\
AS005	& ~00:47:15.9	& ~8.11	& 10.12	& ~21	& ~14	& 21.71	& 0.00184	& 0.04	& 0.74	& 0.95	& 0.78 \\
     	& -20:47:40.5	&      	&      	& ~~8	& ~~8	&      	&        	& 0.09	& 0.09	& 0.09	& 0.10 \\
AS006	& ~00:47:07.6	& ~0.70	& 11.17	& ~23	& ~18	& 22.83	& 0.00120	& 0.04	& 0.72	& 0.97	& 0.75 \\
     	& -20:45:56.0	&      	&      	& ~~7	& ~~7	&   	&        	& 0.23	& 0.09	& 0.09	& 0.09 \\
AS007	& ~00:47:10.3	& ~5.04	& 45.73	& 342	& 160	& 20.32	& 0.00019	& 0.13	& 0.57	& 0.73	& 0.78 \\
     	& -20:48:56.2	&      	&      	& ~77	& ~77	&      	&        	& 0.05	& 0.03	& 0.03	& 0.06 \\
AS008	& ~00:47:10.5	& ~5.26	& 71.56	& 869	& 261	& 20.12	& 0.00004	& 0.15	& 0.47	& 0.81	& 0.57 \\
     	& -20:49:00.1	&      	&      	& ~57	& ~57	&      	&        	& 0.06	& 0.03	& 0.03	& 0.06 \\
AS009	& ~00:47:08.4	& ~0.09	& 40.14	& 323	& ~46	& 20.60	& 0.00006	& 0.12	& 0.54	& 0.80	& 0.67 \\
     	& -20:45:35.3	&      	&      	& ~25	& ~16	&      	&        	& 0.06	& 0.04	& 0.04	& 0.08 \\
AS010	& ~00:47:04.4	& ~3.84	& 45.38	& 350	& ~66	& 20.34	& 0.00010	& 0.14	& 0.51	& 0.71	& 0.72 \\
     	& -20:45:15.8	&      	&      	& ~39	& ~39	&      	&        	& 0.06	& 0.04	& 0.04	& 0.06 \\
\hline
\end{tabular}
\begin{minipage}[]{0.75\textwidth}
{\bf Notes:} 
Columns: 
(1): Identification. 
(2): Equatorial J2000 coordinates. 
(3): Galactocentric distance.
(4): Radius.
(5): Total number of stars ($N_g$) and number of $bright~blue$ stars ($N_{bg}$) inside the group radius.
(6): Total number of stars ($N_d$) and number of $bright~blue$ stars ($N_{bd}$) inside the group radius after field decontamination.
(7): $F606W$ value of the brightest blue star.
(8): Stellar density.
(9): LF slope ($\Gamma$) and its error ($e_{\Gamma}$).
(10): Mean edge length in a MST ($\overline{m}$) and its error ($e_{\overline{m}}$).
(11): Mean separation of the stars in the group ($\overline{s}$) and its error ($e_{\overline{s}}$).
(12): $Q$ parameter and its error ($e_{Q}$).\\
Here we present the first ten groups, the complete table is only available in electronic form.
\end{minipage}
\end{table*}

\begin{figure}
\begin{center}
{\includegraphics[width=0.35\textwidth, trim=0 140 0 140]{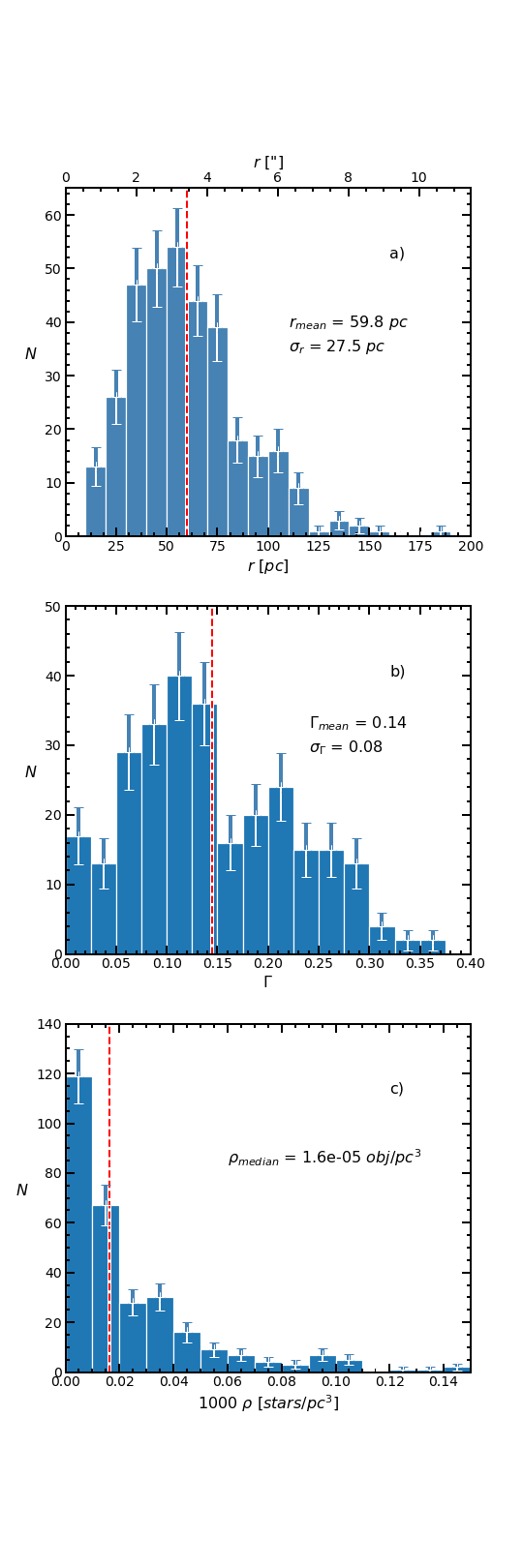}}
\caption{Distribution of sizes (panel a), $\Gamma$ values (panel b) and volumetric densities (panel c) of the the young groups found by the PLC. Error bars arise from the Poisson statistics ($\sqrt N$~). Dashed vertical lines indicate mean or median values.}
\label{distributions}
\end{center}
\end{figure}

\begin{figure*}
\begin{center}
{\includegraphics[width=0.95\textwidth]{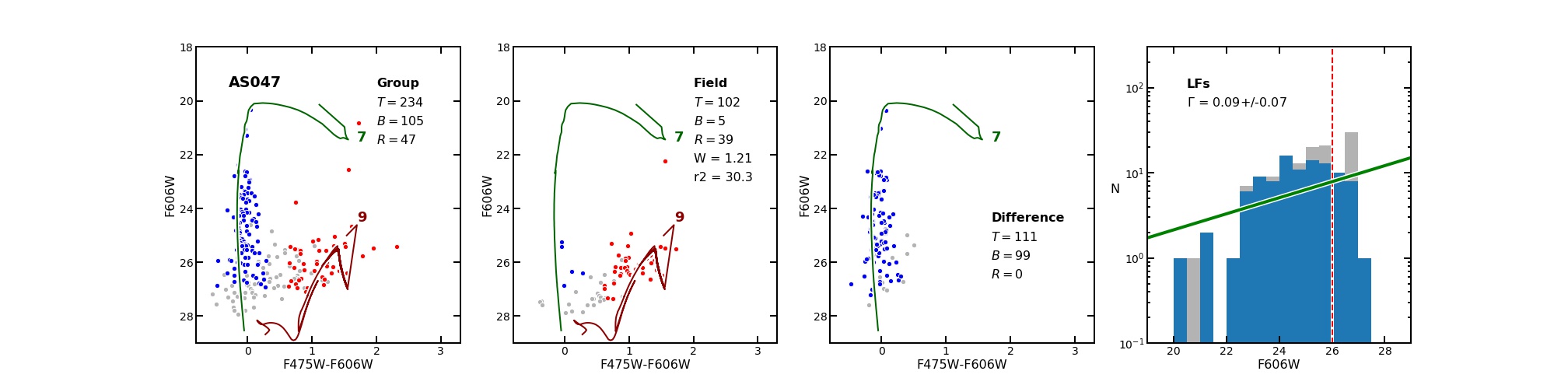}} \\
{\includegraphics[width=0.95\textwidth]{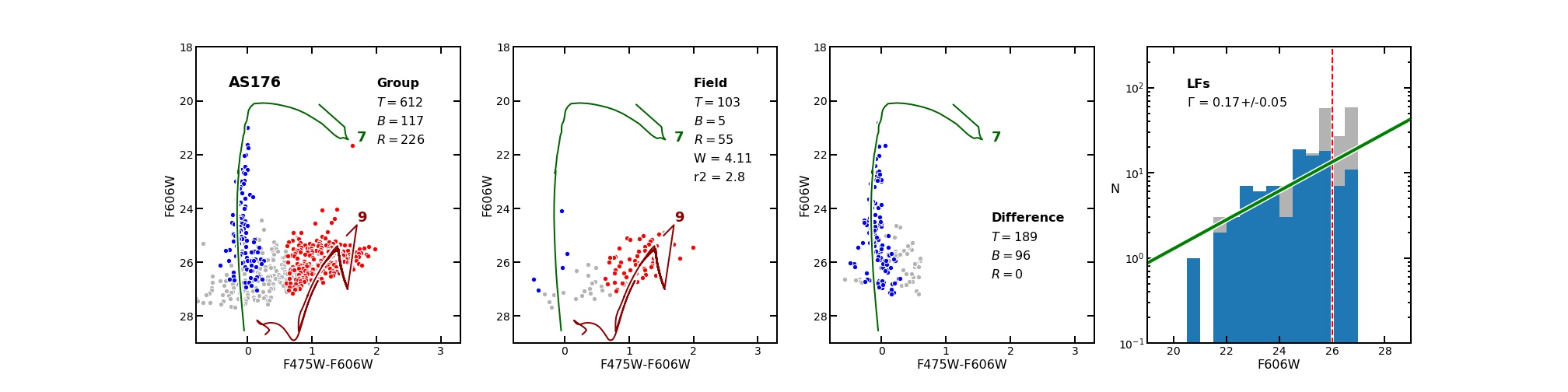}} \\
{\includegraphics[width=0.95\textwidth]{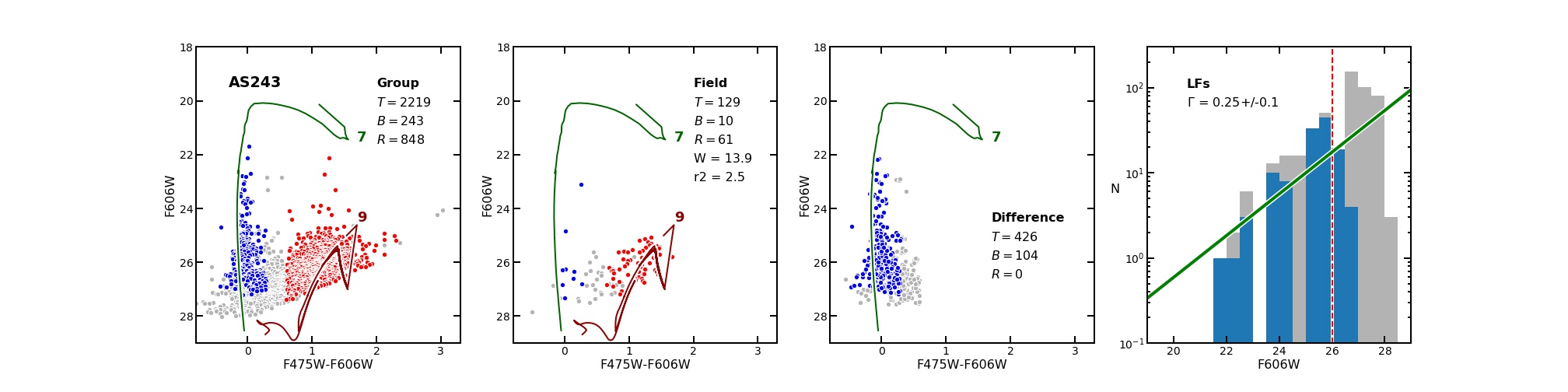}} \\
{\includegraphics[width=0.95\textwidth]{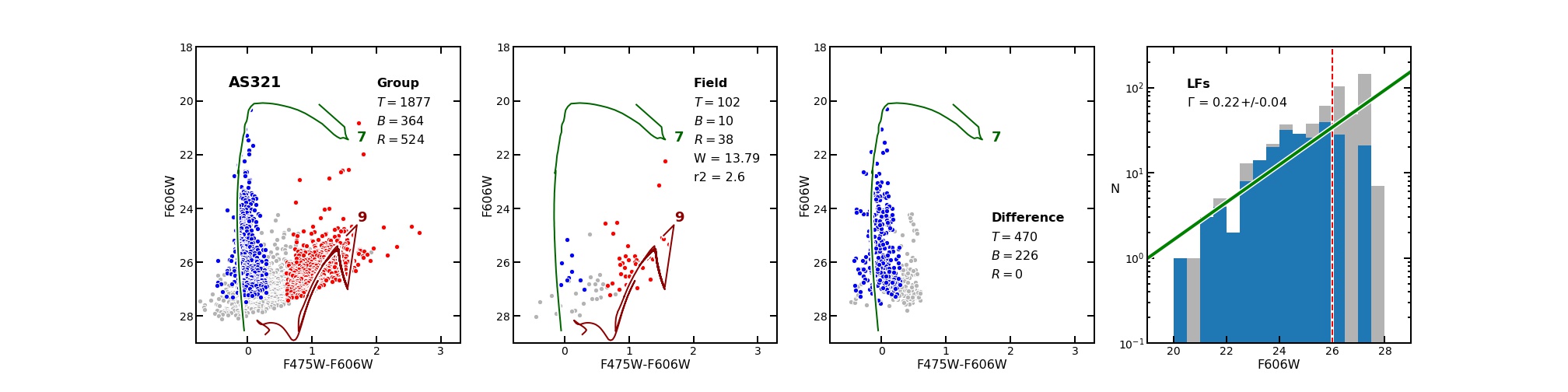}} \\
{\includegraphics[width=0.95\textwidth]{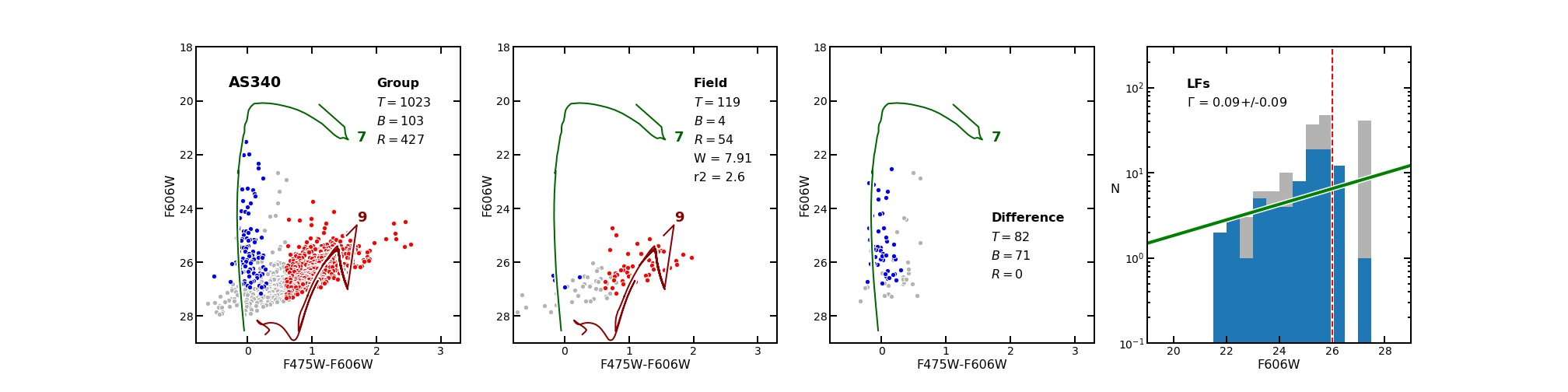}} \\
\caption{CMDs for five selected young stellar groups. Each row presents from left to right: The CMD for the group region, for its corresponding comparison field and the resulting difference. $Blue$ and $red$ symbols indicate selected blue and red objects respectively. The curves are the $10^7$ yr and $10^9$ yr isochrones for solar metallicity ($Z = 0.0152$) from \cite{2014MNRAS.445.4287T} displaced according the adopted distance and reddening for NGC~247. The amount of total ($T$), blue ($B$) and red ($R$) stars are also presented. $W$ and $r2$ indicate respectively the weight and the external radius of the field (see text). In right panels, gray histogram is from all objects remaining after subtraction and blue histogram is only for blue population. The linear fit for brightest objects ($F606W < 26$, dashed line) and the obtained slope value ($\Gamma$) is shown.}
\label{CMDs-LFs}
\end{center}
\end{figure*}

\subsection{Detection of stochastic fluctuations}

We performed numerical simulations using a random distribution of stars with similar densities of those found in the densest areas of the galaxy ($\rho$=0.185 stars~arcsec$^{-2}$) and we ran the PLC over them. These experiments allowed us to estimate the fraction of the PLC stellar groups that should be not real ones, but only stochastic fluctuations of field stars. We did not detect any groups in 10000 experiments with the smallest search radius ($d_s$=0.3 arcsec), with $d_s = 0.7$ arcsec we found only 5 groups in 10000 experiments, and 424 groups with $d_s$=1 arcsec. These fractions are very small strengthening the veracity of the groups detected in the previous section. 

\section{Parameters of the PLC groups}
\label{parameters_sect}

The 339 PLC groups were studied in an homogeneous and systematic way by means of our own numerical code, through which we could estimate the fundamental parameters of the groups.

To estimate the coordinates and the size of the groups, we derived the mean and standard deviation values of the individual positions of the stars identified by the PLC method on each group. Therefore, mean values were adopted as their corresponding centers and twice the standard deviation as the radius ($R$) of a circle enclosing them. The obtained values are presented in the first columns of Table~\ref{tabla1}. The size distribution of the groups is presented in Fig.~\ref{distributions} (panel a) where we can note a large range of radius from 10 to 200 pc, The mode in the distribution is about 30-70 pc and has a mean value of $\sim$ 60 pc. This values are somewhat larger that the ones found in ours previous works for NGC~300 and NGC~253 \citep{2018MNRAS.479..961R,2016A&A...594A..34R}. In the former we found a peak in the size distribution and a mean radius of 25~pc, but it must be considered that in that case we adopted the radius as 1$\sigma$ instead of 2$\sigma$ (as in the present work). Considering $R~=~2~\sigma$, we obtained a mean and a mode of 50~pc, a bit lower that the value found in NGC~247 but still consistent. For NGC~253 we found even lower values, with a mode of 40~pc and a mean radius of 47~pc. In the literature lower values were found too. For example, \cite{2016arXiv160403165D} estimated a mean radius of $\sim$30-35 pc for the young associations in four Local Groups Galaxies (SMC, M31, M33 and NGC 6822) using a friend of friend algorithm. \cite{2013A&A...553A..87D} found a mean size between 62 and 75~pc (radius of $\sim$ 31-38 pc) for the associations in the galaxies NGC~925, NGC~2541, NGC~3351, NGC~3621, NGC~4548 and NGC~5457. On the other hand, a wider range of sizes were found by \citet{2000A&A...357..437B}, who studied the young stellar groups in the galaxies NGC~2717, NGC~1058 and UGC~12732 finding values for the radius mode between 25 and 50 pc, and between 52 and 89 pc for the mean radius. \citet{1991A&A...244...69B} derived a mean radius of 45~pc for the OB associations in SMC, this value is larger than the value found by 
\citet{2016arXiv160403165D} for the same galaxy. In our Galaxy, \citet{1995AstL...21...10M} found still lower values for the OB association sizes in the solar vicinity with a mode value of 15 pc and an average radius of 20 pc.  

Following, we built their corresponding decontaminated CMDs, statistically deleting the field objects located in the CMDs of each group region. This procedure was based on the steps given by \cite{2003A&A...402..549B}. This is, we built the histograms of objects in a three dimensional space of $F606W$, $F434W-F606W$ and $F606W-F814W$ coordinates for each group and for its corresponding comparison field, then both histograms were subtracted. The adopted bin were 0.2 mag wide in magnitude and 0.1 mag in each color. We adopted a circular regions with the corresponding previous computed radius for each stellar group. For the comparison fields we considered an annular shape with an inner and outer radius given by $r1$ and $r2$ times the corresponding radius of the stellar group. To select a clean comparison field we choose $r1 = 1.5$ and we avoided those regions corresponding to detected nearby young groups to that one under study, therefore the choose $r2$ value was a value between 2 and that one corresponding to a comparison field with at least 100 objects. Since the adopted stellar group region and their corresponding field region cover different sky areas, the field histogram was weighted, before the subtraction, by the amount of red stars in both regions. To minimize the amount of unreliable subtractions, this procedure was applied over all the cases were the number of blue objects after the subtraction was larger than the square root of the number of blue object before it. Finally, the decontaminated CMD of each group was generated with random objects following the shape of the resulting three dimensional histogram.

Next, based on the previously obtained decontaminated CMDs, we built their corresponding LFs for all and for blue bright objects and computed the amount of blue objects ($N_b$) of each group. In particular, we also performed a power law fit over the LFs of blue objects and estimated their corresponding $\Gamma$ parameter considering only the brightest region ($F606W < 26$). In Fig.~\ref{CMDs-LFs} we present the $F606W$ vs. $F475W-F606W$ for the circular region of five detected young stellar groups together with those CMDs from their corresponding comparison fields. We also present the obtained subtracted CMDs and the associated LFs for $F606W$ band (see figure caption for details). The corresponding distribution of slopes values is presented in Fig.~\ref{distributions}b, where most of the values spread from 0.05 to 0.30 and it has a bimodal shape with maxima at $\Gamma \sim 0.1$ and $\sim 0.2$. These facts could be understood as different behavior of two stellar populations with differences in age, metallicity and/or the initial mass functions (IMFs). As was already indicated, the general CMDs of NGC~247 (see Fig.~\ref{hess-cmd}) show well defined location for MS stars and blue helium burning stars, therefore the effect of different metallicity would be placed as a second order term. The same can be established for the initial mass distribution effect, since these low slope groups have not preferential location in the galaxy. Therefore, we considered that the bimodal behavior of the $\Gamma$ is mainly due to age differences. In particular, the second peak in the distribution is comparable with the values found in the literature. \citet{2018MNRAS.479..961R} found a mean value of 0.21 for the LF slope in NGC~253. For open cluster in our Galaxy, \citet{2009A&A...504..681K} found a LF slope between 0.2 and 0.3, while \citet{1993AJ....106.1870P} derived a bit larger value of 0.34.       

Additionally, we computed the volumetric density of the stellar groups taking into account the number of decontaminated bright blue objects in an spherical volume with radius $R$. We found then the distribution presented in Fig.~\ref{distributions}c, with a median value of $\rho$ = 1.6 10$^{-5}$ stars pc$^{-3}$. This value was estimated considering only bright objects ($F606W$ < 25) which correspond to approximately B1 spectral type.

In Table~\ref{tabla1} we present the first ten rows of the resulting catalog, in which are listed the properties for each of the 339 young groups. The complete version is only available online. The galactocentric distances in this table were estimated considering the geometric parameters inclination ($i=74^{\circ})$ and position angle ($\gamma=170^{\circ}$) given by \citet{2000A&A...356L..49S}.

\subsection{Correlation with HII regions}

Through a simple visual inspection in the optical HST images, we identified several candidates to HII regions. We characterized them as extended objects with diffuse emission in the studied bands (F475W, F606W and F814W). We detected then 23 candidates, which are listed in Table~\ref{reg_HII}, where we present their coordinates and approximate sizes.

We found a very good correlation between these regions and the PLC groups. We found that 18 HII candidates were related with one or more PLC groups. This fact suggests that most of the HII regions are probably powered by several massive stars, however the HII regions not correlated with a PLC group are probably being stimulated by a few sources that did not pass our PLC criteria ($p$~=~8 stars). 

\begin{table}
\centering 
\caption{Candidates to H{\ttfamily II} regions}  
\label{reg_HII}               
\begin{tabular}{rccc}     
\hline\hline 
ID & $\alpha_{j2000}$ & $\delta_{j2000}$ & $Diameter$ \\
   &   [hh:mm:ss]     &  [dd:mm:ss]      &   ["]      \\
\hline
  1 & 00:47:13.38  & -20:51:40.0 & 3.7\\
  2 & 00:47:13.83  & -20:52:16.6 & 0.9\\
  3 & 00:47:13.29  & -20:52:15.4 & 1.4 \\
  4 & 00:47:13.63  & -20:51:16.0 & 2.2\\
  5 & 00:47:08.81  & -20:50:55.1 & 4.0\\
  6 & 00:47:08.37  & -20:50:30.2 & 1.2\\
  7 & 00:47:08.40  & -20:50:32.2 & 1.5\\
  8 & 00:47:14.37  & -20:50:08.2 & 2.0\\
  9 & 00:47:12.17  & -20:49:21.0 & 4.4\\
 10 & 00:47:12.45  & -20:49:07.3 & 3.2\\
 11 & 00:47:10.21  & -20:48:48.0 & 3.2\\
 12 & 00:47:14.99  & -20:48:38.6 & 3.0\\
 13 & 00:47:15.74  & -20:47:44.7 & 3.8\\
 14 & 00:47:15.92  & -20:47:40.5 & 1.5\\
 15 & 00:47:05.79  & -20:48:05.2 & 2.5\\
 16 & 00:47:17.52  & -20:47:26.6 & 1.0\\
 17 & 00:47:15.78  & -20:47:22.1 & 2.7\\
 18 & 00:47:15.08  & -20:46:07.6 & 5.2\\
 19 & 00:47:07.59  & -20:45:55.3 & 3.0\\
 20 & 00:47:05.10  & -20:45:19.8 & 1.5\\
 21 & 00:47:03.45  & -20:44:45.5 & 0.9\\
 22 & 00:47:11.54  & -20:44:26.3 & 5.5\\
 23 & 00:47:12.43  & -20:46:42.9 & 1.8\\
\hline 
\end{tabular}
\end{table}

\section{Large scale young stellar structures}
\label{largerstructures}

As was mentioned above, young stellar groups exist in a large variety of scales, from star clusters to stellar complexes, or even larger structures. Several authors \citep[e.g.][]{2018PASP..130g2001G, 2007MNRAS.379.1302B} proposed that star formation occurs in a continuous fashion, characterized by an hierarchical pattern; where the star cluster and association are the denser inner part.  

\begin{figure}
\includegraphics[width=0.5\textwidth, trim=0 80 0 80]{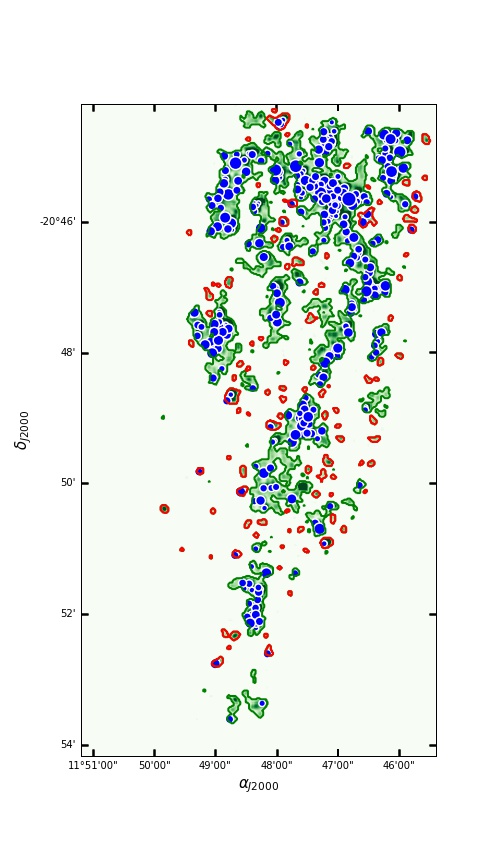} \\
\caption{Kernel density map of brightest blue stars in the studied region of NGC~247 obtained using a Gaussian kernel with a 2 arcsec bandwidth. Green and red levels indicate regions with and without hierarchical structure respectively. Blue circles present the location of identified young stellar groups using PLC.}
\label{KDE}
\end{figure}

To identify larger young stellar structures, we constructed a surface stellar density map over the bright and blue stars of the NGC~247 studied region (see Sect.~\ref{populations}). This was performed using the kernel density estimation (KDE) method. This is, we convolved the stellar distribution with a Gaussian kernel. This method requires choosing a characteristic bandwidth for the kernel, this selection was done with an empirical approach. We first employed the "stratified K-folds cross-validation" and a "Grid Search" method to select the best parameters for an algorithm, in brief, a range of bandwidth values were tested over a sample of the data, and the one that produced a minor error was chosen. In our case, the bandwidth values ranged from 1 arcsec to 10 arcsec and a 20 fold test. These steps led to a minimum error bandwidth of 3 arcsec. Following, we built the corresponding stellar density map using this bandwidth value and lower ones, checking in all these maps that the main structures were still preserved and that these structures complemented those ones found using the PLC method. We finally adopted a bandwidth of 2 arcsec. The final adopted stellar density map of the bright blue stars is presented in Fig.~\ref{KDE}. The minimum error bandwidth value estimation and KDE map construction were performed using respectively the "GridSearchCV" and "KernelDensity" tasks of Python "Scikit-learn" library (see \citealt{scikit-learn} or https://scikit-learn.org for more details). 

Several features were evident in the obtained stellar density map. In particular the main overdensity at the galactic center, the NE-SW bar direction and the two main spiral arms. In addition to a large number of smaller structures, both inside and outside the main structures. 
In fact, the adopted 2 arcsec bandwidth corresponds to $\sim$35 pc (standard deviation) at the NGC~247 adopted distance, and this size was the limit of resolved structures in the KDE map.

\begin{figure*}
\begin{tabular}{cc}
\includegraphics[width=0.43\textwidth, trim=70 400 70 
400]{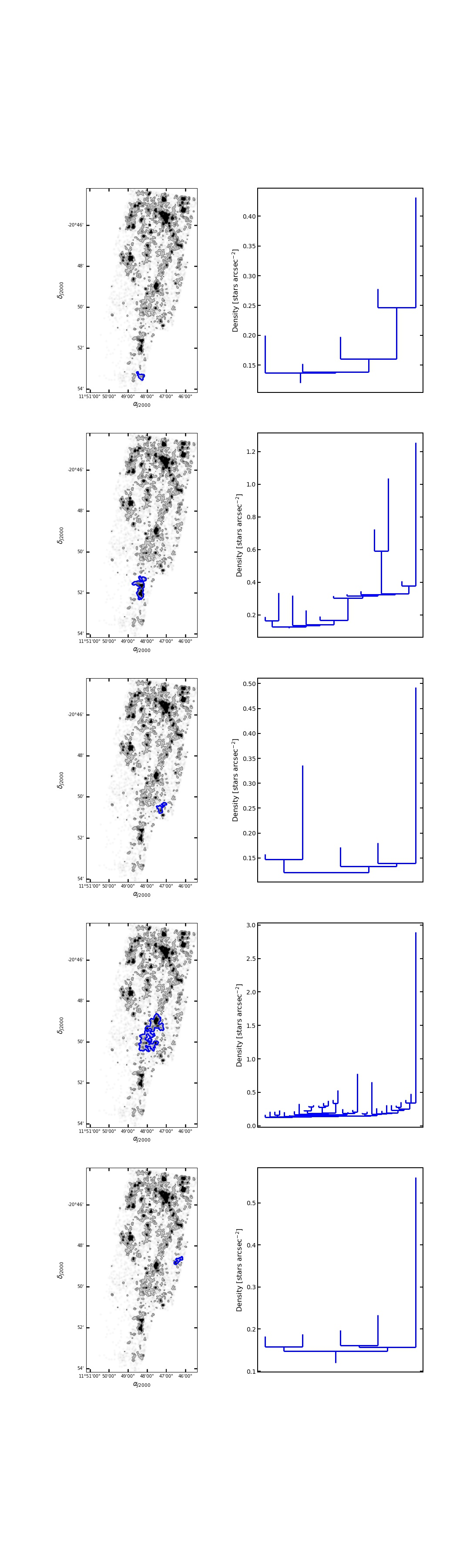} &
\includegraphics[width=0.43\textwidth, trim=70 400 70 
400]{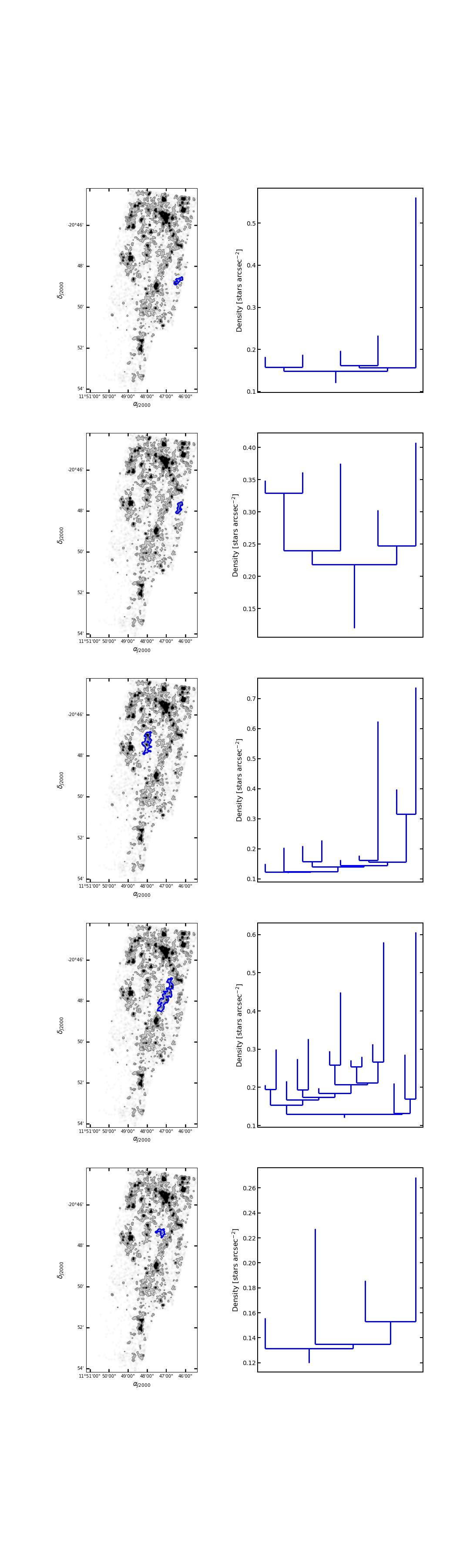} \\
\end{tabular}
\caption{Dendrograms of 15 main NGC~247 structures. Left panels identify in blue each structure over the KDE, whereas right panels present the corresponding dendrogram.}
\label{dendrogram}
\end{figure*}

To identify structures in the young stellar population of NGC~247, we build the corresponding dendrograms of the KDE map using the Core algorithm implemented in the "astrodenro" library in Python (see https://dendrograms.readthedocs.io/en/stable/ for details). 
 Basically, the generated dendrogram obtained from the KDE map is a "tree" type graphic diagram representing the skeleton of the map. In this representation, the "leaves" indicate the density peaks of the map and their joining represent the valleys (see \citealt{2008ApJ...679.1338R} for more details). The diagram follows until a minimum density threshold ($\rho_{min}$) is reached. In order to avoid in the diagram those peaks from noise fluctuation in the density map, it is necessary to indicate also a minimum acceptable fluctuation ($\Delta \rho_{min}$). After several tests with different pair of these parameters, we adopted $\rho_{min} = 0.12~stars~ arcsec^{-2}$ and $\Delta \rho_{min} = 0.01~stars~ arcsec^{-2}$ since they produced clear dendrograms.
This procedure allowed to identify two kinds of overdensities (above $\rho_{min}$): 28 with internal substructures (green ones in Fig.~\ref{KDE}) and 102 without substructures (red ones in Fig.~\ref{KDE}). The corresponding diagrams for the 15 most important structures (with more than 4 decendents in each one) are presented in Fig.~\ref{dendrogram}, where some relevant parameters of each remarked structure are also indicated. These large scale structures complemented the small scale ones already found using the PLC method (see Sect.~\ref{plc_sect}). Only 18 young stellar groups found by PLC are placed at overdensities without substructures.
This represent a small percentage ($\sim 5 \%$), similar percentages were found in other galaxies \citep[e.g][]{2017MNRAS.468..509G,1995AJ....110.2757E}. These few groups could be formed by different initial condition in the ISM, or perhaps they are older and their larger associated structures are too loose to be detected as over densities in the density map. 

\renewcommand{\thefigure}{\arabic{figure} (Cont.)}
\addtocounter{figure}{-1}
\begin{figure}
\begin{tabular}{cc}
\includegraphics[width=0.43\textwidth, trim=70 400 70 
400]{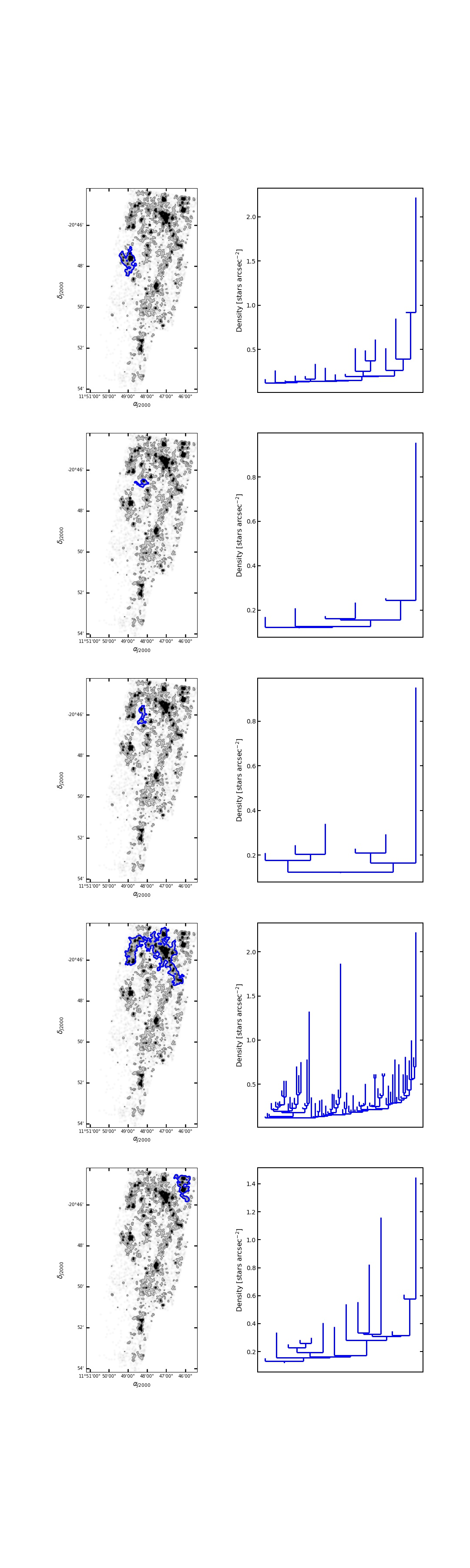}
\end{tabular}
\caption{}
\end{figure}
\renewcommand{\thefigure}{\arabic{figure}}

\section{Morphological analysis}
\label{morphology}

Figures~\ref{KDE} and \ref{dendrogram} reveal the presence of a hierarchical organization of the young stellar population, in which a structure with low stellar density may contain one or more substructures with higher stellar densities inside it. This hierarchical structure is considered typical in several star-forming regions and galaxies  \citep[e.g][]{2018PASP..130g2001G, 2018ApJ...858...31S, 2018MNRAS.479..961R, 2014MNRAS.442.3711G}, and could be inherited from their parent molecular clouds, which present a self similar or fractal distribution \citep{1999ApJ...527..266E}. In fact, this fractal behavior is consistent with a scenario of star formation regulated by turbulence and self-gravity of the interstellar medium (ISM), in which the molecular cloud is fragmented in successive smaller clouds \citep{1996ApJ...466..802E}.

\begin{figure}
\begin{center}
\includegraphics[width=0.35\textwidth, trim=10 100 10 100]{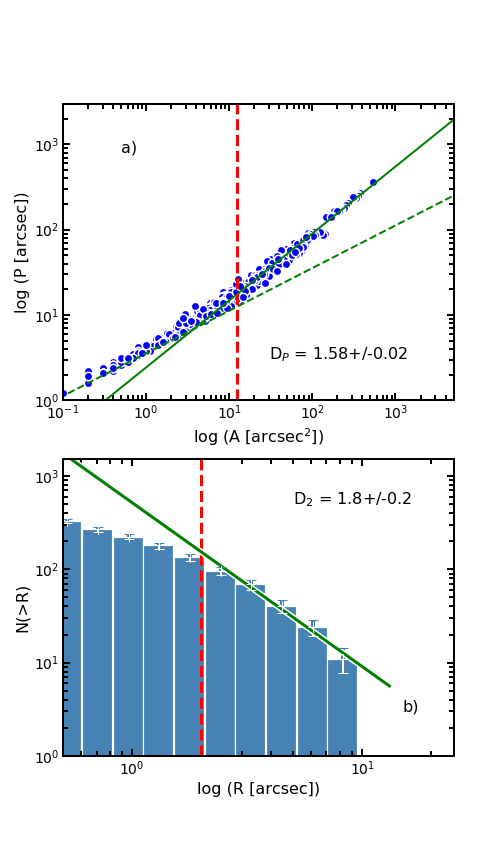} \\
\caption{Perimeter-area relation (panel a) and cumulative size histogram (panel b) for the young stellar structures detected in the density map. The vertical dashed lines indicate the resolution limit of the KDE map and solid line are the best fits following the behavior of large structures. Fractal dimension values obtained in each case are indicated. In panel a) the dashed sloped line shows the relation for a circle ($P \propto A^{0.5}$). In panel b) errors are from $\sqrt N$~.}
\label{fractal}
\end{center}
\end{figure}

We investigated the properties of the clustering in the young structures detected in Sects.~\ref{S_search} and \ref{largerstructures}. This is, we studied their corresponding fractal behavior using different methodologies. In the case of large scale structures, we estimated the fractal dimension \citep[$D_2$,][]{mandelbrot1982fractal} through the perimeter ($P$) area ($A$) relation and the cumulative size ($R$) distribution. In the former case the relation $P \propto A^{D_2/2}$ \citep{1991ApJ...378..186F, 2018ApJ...858...31S} is verified, whereas in the latter, the following relation $dN/dlog(R) \propto R^{-D}$ \citep{mandelbrot1982fractal, 1996ApJ...471..816E} is valid. The obtained values in each case were  $1.58 \pm 0.02$ and $1.8 \pm 0.2$.
These values are close to those reported for other star forming regions, as well as for several galaxies, and agree with a scenario of hierarchical stellar formation in which newly born stars follow the gas distribution \citep{2017ApJ...835..171S,2018ApJ...858...31S}.
In Fig. ~\ref{fractal} we show both relations with their corresponding fractal dimension fits. In panel a, we can see that the structures with sizes under the resolution limit follow the perimeter area relation for a circle.

\begin{figure}
\begin{center}
\includegraphics[width=0.3\textwidth, trim=10 90 10 
90]{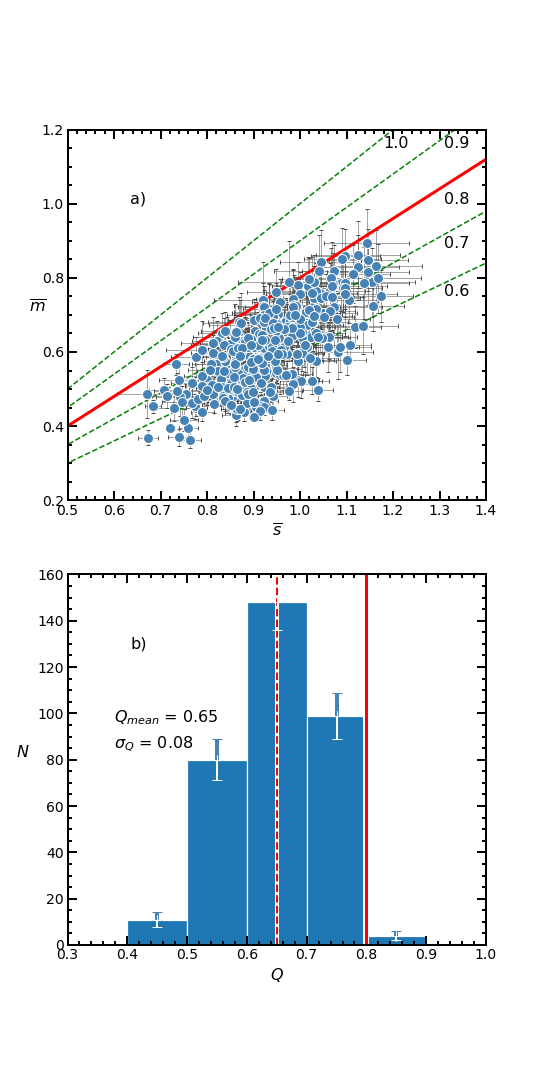}
\caption{{\bf a)} Trend of $\overline{m}$ vs. $\overline{s}$ for the 339 PLC groups. Lines indicate the location for different $Q$ values. Error bars are also presented. {\bf b)} Distribution of $Q$ values. Red line indicates $Q = 0.8$. Error bars are computed as $\sqrt N$~.}
\label{Q}
\end{center}
\end{figure}

In the case of small structures, we computed the $Q$ parameter introduced by \citet{2004MNRAS.348..589C} to distinguish between a smooth large scale radial density gradient and multiscale fractal clustering. The $Q$ parameter is defined as the ratio between two parameters: $\overline{m}$, which is the mean edge length in a minimum spanning tree (MST), normalized by the cluster area, and $\overline{s}$, which is the mean separation of the points, normalized by the cluster ratio. This is:  \\

$\displaystyle\overline{m}=\frac{1}{(AN)^{1/2}}$ $\displaystyle\sum_{i=1}^{N-1}m_i$

$\displaystyle\bar{s}=\frac{2}{N(N-1)R_{sc}}\displaystyle\sum_{i=1}^{N-1}\displaystyle\sum_{j=i+1}^{N}\left|\vec{r_i}-\vec{r_j} \right|$,
 
\noindent where $N$ is the number of objects connected by the PLC, $A$ is the area of the smallest circle that contains all the objects, $R_{sc}$ is its corresponding radius and $r_i$ is the position of the $i$ object. It was found that $R_{sc} \sim 0.8 R$, where $R$ is the adopted radius for the stellar groups computed in Sect~\ref{parameters_sect} and presented in Table~\ref{tabla1}. 

To estimate $\overline{m}$, we constructed the MST of each stellar group, which is the shortest possible network that connects all the stars in the cluster and there are no closed loops. The MST was built using the "minimum\_spanning\_tree" task at Python "scipy.sparse.csgraph" library. 

\citet{2004MNRAS.348..589C} obtained for artificial cluster that values of $Q$ greater than 0.8 correspond to a central concentrated distribution, while smaller values agree with fractal structures. In this last case, $Q$ values decreasing in the range 0.8~-~0.45 are related to a fractal dimension ($D_2$) decreasing from 2 to 0.5.

We computed then $\overline{m}$, $\overline{s}$ and $Q$ for all groups detected in Sect.~\ref{plc_sect}. We also evaluated their corresponding errors obtaining these values several times ($\sim$ 100 experiments) on each stellar group using the "bootstrap" method \citep{2015MNRAS.448.2504G} and computing the standard deviations of the different values computed on each experiment. The obtained results are in the last columns of Table~\ref{tabla1} and represented in Fig.~\ref{Q}. In particular the computed errors values were $e_{\overline{m}} \sim$ 0.08, $e_{\overline{s}} \sim$ 0.08 and $e_Q \sim$ 0.10. From Fig.~\ref{Q} we noticed that most of the $Q$ values were lower than 0.8 and only a few were above this value but they could be explained by the uncertainty of these values. Therefore it was possible to conclude that all the identified young stellar groups reveal a fractal structure. This fact could suggest that the young groups detected in the present work are very young and still show the distribution of their parent molecular clouds \citep{2018MNRAS.481.1016G,2010LNEA....4....1B}.

\section{Conclusions}
\label{conclusions}

We performed an analysis of the young stellar population in the Sculptor group galaxy NGC~247. 
Through the PLC method we found 339 young stellar groups that could be open clusters, stellar associations, or a bit larger young groups.

We derived a catalog contains the main characterizes of each group.
For these groups we derived a mean radius of 60 pc and a mode in the size distribution between 30 and 70 pc. This values are a little larger than the radii found in the literature for most galaxies including our Galaxy.
We also derived a mean density of 1.6 10$^{-5} stars~pc^{-3}$ computing stars brighter than $B1$. The LF slope distribution shows a bimodal shape
with maximums at 0.1 and 0.2.
This could be interpreted as the presence of two stellar population with different characterizes. 
The first value, which corresponds to the higher one in the distribution, represents flatter LF slopes. This population might be younger and so it would have more bright stars, 
causing the LF slope to become flatter. The second value is consistent with the one found for NGC~253 and The Milky Way.
We detected a list of HII region candidates finding an excellent correlation between them and the PLC groups.

On the other hand, we studied the internal structure of each of the 339 PLC groups by means of the $Q$ parameter.
We obtained that all of the sample presents sub-clumpings. 
This fact suggests that the studied population is very young and the groups still preserve the same structure of the molecular clouds from which they form. 
This idea is supported by the flat values in the LF slope. 
The young groups are expected to evolve from a internal distribution with sub-structures to a more uniform one with time.

We detected larger young stellar structures through the blue stars density map. For these structures a high degree of clustering is revealed by the dendrograms
showed in Fig.~\ref{dendrogram}. By means of the perimeter-area relation and the cumulative size distribution for these structures we obtained fractal dimensions
of 1.58$\pm$0.02 and 1.8$\pm$0.2 respectively. These values are consistent with the ones obtained in other star forming regions and galaxies, and are consistent with a scenario of hierarchical star formation regulated by supersonic turbulence and self gravity. 

We found a small percentage ($\sim 5\%$) of the young star groups that do not seems to be part of the global hierarchical structure observed. This small percentage was also found in others galaxies and could be pointing different initial conditions of the ISM when these groups form. Another possibility is that these groups are older and therefore
their associated major structures have evolve in a loose population difficult to detect from background stars.

\begin{acknowledgements} 

We thank the referee for helpful comments and constructive suggestions that helped to improve this paper.
MJR and GB acknowledge support from CONICET (PIP 112-201701-00055). MJR is a fellow of CONICET. This work was based on observations made with the NASA/ESA Hubble Space Telescope, and obtained from the Hubble Legacy Archive, which is a collaboration between the Space Telescope Science Institute (STScI/NASA), the Space Telescope European Coordinating Facility (ST-ECF/ESA) and the Canadian Astronomy Data Centre (CADC/NRC/CSA). Some of the data presented in this paper were obtained from the Mikulski Archive for Space Telescopes (MAST). STScI is operated by the Association of Universities for Research in Astronomy, Inc., under NASA contract NAS5-26555. Support for MAST for non-HST data is provided by the NASA Office of Space Science via grant NNX09AF08G and by other grants and contracts. This research has made use of "Aladin sky atlas" developed at CDS, Strasbourg Observatory, France. We also made use of astrodendro, a Python package to compute dendrograms of Astronomical data (http://www.dendrograms.org/)
 
\end{acknowledgements}

\bibliographystyle{aa} 
\bibliography{references} 

\begin{thebibliography}{49}
\expandafter\ifx\csname natexlab\endcsname\relax\def\natexlab#1{#1}\fi

\bibitem[{{Bastian} {et~al.}(2007){Bastian}, {Ercolano}, {Gieles},
  {Rosolowsky}, {Scheepmaker}, {Gutermuth}, \& {Efremov}}]{2007MNRAS.379.1302B}
{Bastian}, N., {Ercolano}, B., {Gieles}, M., {et~al.} 2007, \mnras, 379, 1302

\bibitem[{{Battinelli}(1991)}]{1991A&A...244...69B}
{Battinelli}, P. 1991, \aap, 244, 69

\bibitem[{{Battinelli} {et~al.}(2000){Battinelli}, {Capuzzo-Dolcetta}, {Hodge},
  {Vicari}, \& {Wyder}}]{2000A&A...357..437B}
{Battinelli}, P., {Capuzzo-Dolcetta}, R., {Hodge}, P.~W., {Vicari}, A., \&
  {Wyder}, T.~K. 2000, \aap, 357, 437

\bibitem[{{Baume} {et~al.}(2003){Baume}, {V{\'a}zquez}, {Carraro}, \&
  {Feinstein}}]{2003A&A...402..549B}
{Baume}, G., {V{\'a}zquez}, R.~A., {Carraro}, G., \& {Feinstein}, A. 2003,
  \aap, 402, 549

\bibitem[{{Blaauw}(1964)}]{1964ARA&A...2..213B}
{Blaauw}, A. 1964, ARA\&A, 2, 213

\bibitem[{{Carignan} \& {Puche}(1990)}]{1990AJ....100..641C}
{Carignan}, C. \& {Puche}, D. 1990, AJ, 100, 641

\bibitem[{{Cartwright} \& {Whitworth}(2004)}]{2004MNRAS.348..589C}
{Cartwright}, A. \& {Whitworth}, A.~P. 2004, \mnras, 348, 589

\bibitem[{{Dalcanton} {et~al.}(2008){Dalcanton}, {Williams}, \& {ANGST
  Collaboration}}]{2008ASSP....5..115D}
{Dalcanton}, J., {Williams}, B., \& {ANGST Collaboration}. 2008, Astrophysics
  and Space Science Proceedings, 5, 115

\bibitem[{{Davidge}(2006)}]{2006ApJ...641..822D}
{Davidge}, T.~J. 2006, ApJ, 641, 822

\bibitem[{{Dolphin}(2000)}]{2000PASP..112.1383D}
{Dolphin}, A.~E. 2000, \pasp, 112, 1383

\bibitem[{{Drazinos} {et~al.}(2016){Drazinos}, {Karampelas}, {Kontizas},
  {Kontizas}, {Dapergolas}, {Livanou}, \&
  {Bellas-Velidis}}]{2016arXiv160403165D}
{Drazinos}, P., {Karampelas}, A., {Kontizas}, E., {et~al.} 2016, arXiv e-prints
  [\eprint[arXiv]{1604.03165}]

\bibitem[{{Drazinos} {et~al.}(2013){Drazinos}, {Kontizas}, {Karampelas},
  {Kontizas}, \& {Dapergolas}}]{2013A&A...553A..87D}
{Drazinos}, P., {Kontizas}, E., {Karampelas}, A., {Kontizas}, M., \&
  {Dapergolas}, A. 2013, \aap, 553, A87

\bibitem[{{Efremov}(1995)}]{1995AJ....110.2757E}
{Efremov}, Y.~N. 1995, \aj, 110, 2757

\bibitem[{{Efremov} \& {Elmegreen}(1998)}]{1998MNRAS.299..588E}
{Efremov}, Y.~N. \& {Elmegreen}, B.~G. 1998, MNRAS, 299, 588

\bibitem[{{Elmegreen}(1999)}]{1999ApJ...527..266E}
{Elmegreen}, B.~G. 1999, ApJ, 527, 266

\bibitem[{{Elmegreen} \& {Efremov}(1996)}]{1996ApJ...466..802E}
{Elmegreen}, B.~G. \& {Efremov}, Y.~N. 1996, ApJ, 466, 802

\bibitem[{{Elmegreen} \& {Falgarone}(1996)}]{1996ApJ...471..816E}
{Elmegreen}, B.~G. \& {Falgarone}, E. 1996, \apj, 471, 816

\bibitem[{{Elmegreen} \& {Salzer}(1999)}]{1999AJ....117..764E}
{Elmegreen}, D.~M. \& {Salzer}, J.~J. 1999, \aj, 117, 764

\bibitem[{{Falgarone} {et~al.}(1991){Falgarone}, {Phillips}, \&
  {Walker}}]{1991ApJ...378..186F}
{Falgarone}, E., {Phillips}, T.~G., \& {Walker}, C.~K. 1991, \apj, 378, 186

\bibitem[{{Feinstein} {et~al.}(2019){Feinstein}, {Baume}, {Rodriguez}, \&
  {Vergne}}]{Feinstein}
{Feinstein}, C., {Baume}, G., {Rodriguez}, J., \& {Vergne}, M. 2019, arXiv
  e-prints [\eprint[arXiv]{1903.10989}]

\bibitem[{{Ferguson} {et~al.}(1996){Ferguson}, {Wyse}, {Gallagher}, \&
  {Hunter}}]{1996AJ....111.2265F}
{Ferguson}, A.~M.~N., {Wyse}, R.~F.~G., {Gallagher}, III, J.~S., \& {Hunter},
  D.~A. 1996, AJ, 111, 2265

\bibitem[{{Garc{\'{\i}}a-Varela} {et~al.}(2008){Garc{\'{\i}}a-Varela},
  {Pietrzy{\'n}ski}, {Gieren}, {Udalski}, {Soszy{\'n}ski}, {Walker},
  {Bresolin}, {Kudritzki}, {Szewczyk}, {Szyma{\'n}ski}, {Kubiak}, \&
  {Wyrzykowski}}]{2008AJ....136.1770G}
{Garc{\'{\i}}a-Varela}, A., {Pietrzy{\'n}ski}, G., {Gieren}, W., {et~al.} 2008,
  \aj, 136, 1770

\bibitem[{{Girardi} {et~al.}(2012){Girardi}, {Barbieri}, {Groenewegen},
  {Marigo}, {Bressan}, {Rocha-Pinto}, {Santiago}, {Camargo}, \& {da
  Costa}}]{Trigal}
{Girardi}, L., {Barbieri}, M., {Groenewegen}, M.~A.~T., {et~al.} 2012,
  Astrophysics and Space Science Proceedings, 26, 165

\bibitem[{{Gouliermis}(2018)}]{2018PASP..130g2001G}
{Gouliermis}, D.~A. 2018, \pasp, 130, 072001

\bibitem[{{Gouliermis} {et~al.}(2017){Gouliermis}, {Elmegreen}, {Elmegreen},
  {Calzetti}, {Cignoni}, {Gallagher}, {Kennicutt}, {Klessen}, {Sabbi},
  {Thilker}, {Ubeda}, {Aloisi}, {Adamo}, {Cook}, {Dale}, {Grasha}, {Grebel},
  {Johnson}, {Sacchi}, {Shabani}, {Smith}, \& {Wofford}}]{2017MNRAS.468..509G}
{Gouliermis}, D.~A., {Elmegreen}, B.~G., {Elmegreen}, D.~M., {et~al.} 2017,
  \mnras, 468, 509

\bibitem[{{Grasha} {et~al.}(2018){Grasha}, {Calzetti}, {Bittle}, {Johnson},
  {Donovan Meyer}, {Kennicutt}, {Elmegreen}, {Adamo}, {Krumholz}, {Fumagalli},
  {Grebel}, {Gouliermis}, {Cook}, {Gallagher}, {Aloisi}, {Dale}, {Linden},
  {Sacchi}, {Thilker}, {Walterbos}, {Messa}, {Wofford}, \&
  {Smith}}]{2018MNRAS.481.1016G}
{Grasha}, K., {Calzetti}, D., {Bittle}, L., {et~al.} 2018, \mnras, 481, 1016

\bibitem[{{Gregorio-Hetem} {et~al.}(2015){Gregorio-Hetem}, {Hetem},
  {Santos-Silva}, \& {Fernandes}}]{2015MNRAS.448.2504G}
{Gregorio-Hetem}, J., {Hetem}, A., {Santos-Silva}, T., \& {Fernandes}, B. 2015,
  \mnras, 448, 2504

\bibitem[{{Gusev}(2014)}]{2014MNRAS.442.3711G}
{Gusev}, A.~S. 2014, \mnras, 442, 3711

\bibitem[{{Kharchenko} {et~al.}(2009){Kharchenko}, {Piskunov}, {R{\"o}ser},
  {Schilbach}, {Scholz}, \& {Zinnecker}}]{2009A&A...504..681K}
{Kharchenko}, N.~V., {Piskunov}, A.~E., {R{\"o}ser}, S., {et~al.} 2009, A\&A,
  504, 681

\bibitem[{{Kiss} \& {Bedding}(2005)}]{Kiss}
{Kiss}, L.~L. \& {Bedding}, T.~R. 2005, \mnras, 358, 883

\bibitem[{{Konstantopoulos} {et~al.}(2013){Konstantopoulos}, {Smith}, {Adamo},
  {Silva-Villa}, {Gallagher}, {Bastian}, {Ryon}, {Westmoquette}, {Zackrisson},
  {Larsen}, {Weisz}, \& {Charlton}}]{2013AJ....145..137K}
{Konstantopoulos}, I.~S., {Smith}, L.~J., {Adamo}, A., {et~al.} 2013, \aj, 145,
  137

\bibitem[{{Lada} \& {Lada}(2003)}]{2003ARA&A..41...57L}
{Lada}, C.~J. \& {Lada}, E.~A. 2003, \araa, 41, 57

\bibitem[{{Larsen}(1999)}]{1999A&AS..139..393L}
{Larsen}, S.~S. 1999, \aaps, 139, 393

\bibitem[{Mandelbrot(1982)}]{mandelbrot1982fractal}
Mandelbrot, B.~B. 1982, The fractal geometry of nature, Vol.~1 (WH freeman New
  York)

\bibitem[{{Melnik} \& {Efremov}(1995)}]{1995AstL...21...10M}
{Melnik}, A.~M. \& {Efremov}, Y.~N. 1995, Astronomy Letters, 21, 10

\bibitem[{{O'Donnell}(1994)}]{1994ApJ...437..262O}
{O'Donnell}, J.~E. 1994, \apj, 437, 262

\bibitem[{{Olsen} {et~al.}(2004){Olsen}, {Miller}, {Suntzeff}, {Schommer}, \&
  {Bright}}]{2004AJ....127.2674O}
{Olsen}, K.~A.~G., {Miller}, B.~W., {Suntzeff}, N.~B., {Schommer}, R.~A., \&
  {Bright}, J. 2004, AJ, 127, 2674

\bibitem[{Pedregosa {et~al.}(2011)Pedregosa, Varoquaux, Gramfort, Michel,
  Thirion, Grisel, Blondel, Prettenhofer, Weiss, Dubourg, Vanderplas, Passos,
  Cournapeau, Brucher, Perrot, \& Duchesnay}]{scikit-learn}
Pedregosa, F., Varoquaux, G., Gramfort, A., {et~al.} 2011, Journal of Machine
  Learning Research, 12, 2825

\bibitem[{{Phelps} \& {Janes}(1993)}]{1993AJ....106.1870P}
{Phelps}, R.~L. \& {Janes}, K.~A. 1993, AJ, 106, 1870

\bibitem[{{Rodr{\'{\i}}guez} {et~al.}(2016){Rodr{\'{\i}}guez}, {Baume}, \&
  {Feinstein}}]{2016A&A...594A..34R}
{Rodr{\'{\i}}guez}, M.~J., {Baume}, G., \& {Feinstein}, C. 2016, \aap, 594, A34

\bibitem[{{Rodr{\'{\i}}guez} {et~al.}(2018){Rodr{\'{\i}}guez}, {Baume}, \&
  {Feinstein}}]{2018MNRAS.479..961R}
{Rodr{\'{\i}}guez}, M.~J., {Baume}, G., \& {Feinstein}, C. 2018, \mnras, 479,
  961

\bibitem[{{Rosolowsky} {et~al.}(2008){Rosolowsky}, {Pineda}, {Kauffmann}, \&
  {Goodman}}]{2008ApJ...679.1338R}
{Rosolowsky}, E.~W., {Pineda}, J.~E., {Kauffmann}, J., \& {Goodman}, A.~A.
  2008, \apj, 679, 1338

\bibitem[{{S{\'a}nchez} \& {Alfaro}(2010)}]{2010LNEA....4....1B}
{S{\'a}nchez}, N. \& {Alfaro}, E.~J. 2010, Lecture Notes and Essays in
  Astrophysics, 4, 1

\bibitem[{{Schaap} {et~al.}(2000){Schaap}, {Sancisi}, \&
  {Swaters}}]{2000A&A...356L..49S}
{Schaap}, W.~E., {Sancisi}, R., \& {Swaters}, R.~A. 2000, A\&A, 356, L49

\bibitem[{{Schlafly} \& {Finkbeiner}(2011)}]{2011ApJ...737..103S}
{Schlafly}, E.~F. \& {Finkbeiner}, D.~P. 2011, \apj, 737, 103

\bibitem[{{Sun} {et~al.}(2018){Sun}, {de Grijs}, {Cioni}, {Rubele},
  {Subramanian}, {van Loon}, {Bekki}, {Bell}, {Ivanov}, {Marconi}, {Muraveva},
  {Oliveira}, \& {Ripepi}}]{2018ApJ...858...31S}
{Sun}, N.-C., {de Grijs}, R., {Cioni}, M.-R.~L., {et~al.} 2018, \apj, 858, 31

\bibitem[{{Sun} {et~al.}(2017){Sun}, {de Grijs}, {Subramanian}, {Cioni},
  {Rubele}, {Bekki}, {Ivanov}, {Piatti}, \& {Ripepi}}]{2017ApJ...835..171S}
{Sun}, N.-C., {de Grijs}, R., {Subramanian}, S., {et~al.} 2017, \apj, 835, 171

\bibitem[{{Tang} {et~al.}(2014){Tang}, {Bressan}, {Rosenfield}, {Slemer},
  {Marigo}, {Girardi}, \& {Bianchi}}]{2014MNRAS.445.4287T}
{Tang}, J., {Bressan}, A., {Rosenfield}, P., {et~al.} 2014, \mnras, 445, 4287

\bibitem[{{Tao} {et~al.}(2012){Tao}, {Feng}, {Kaaret}, {Gris{\'e}}, \&
  {Jin}}]{2012ApJ...758...85T}
{Tao}, L., {Feng}, H., {Kaaret}, P., {Gris{\'e}}, F., \& {Jin}, J. 2012, ApJ,
  758, 85

\end{thebibliography}

\end{document}